\documentclass{sig-alternate}

\usepackage{url,graphicx}

\begin{document}
\conferenceinfo{HT'09,} {June 29--July 1, 2009, Torino, Italy.} 

\CopyrightYear{2009}

\crdata{978-1-60558-486-7/09/06} 

\title{Statistical Properties of Inter-arrival Times Distribution \\ in Social Tagging Systems}
%
%
% Put no more than the first THREE authors in the \author command
% NOTE: All authors should be on the first page. For instructions
% for more than 3 authors, see:
% http://www.acm.org/sigs/pubs/proceed/sigfaq.htm#a18
%
\numberofauthors{4}
\author{
\alignauthor Andrea Capocci\\
\affaddr{University ''La Sapienza'' of Rome, Italy}\\
\affaddr{P.le Aldo Moro, 5, Rome, Italy}\\
%    \email{andrea.capocci@gmail.com}\\
%
%
\alignauthor Andrea Baldassarri\\
\affaddr{University ``La Sapienza'' of Rome, Italy}\\
\affaddr{P.le Aldo Moro, 5, Rome, Italy}
\and
\alignauthor Vito D.P. Servedio\\
\affaddr{University ``La Sapienza'' of Rome, Italy}\\
\affaddr{P.le Aldo Moro, 5, Rome, Italy}
\alignauthor Vittorio Loreto\\
\affaddr{University ``La Sapienza'' of Rome, Italy}\\
\affaddr{P.le Aldo Moro, 5, Rome, Italy}
\affaddr{ISI Foundation}\\
\affaddr{Villa Gualino, Turin, Italy}
}

\maketitle
\begin{abstract}
Folksonomies provide a rich source of data to study social patterns
taking place on the World Wide Web. Here we study the temporal
patterns of users' tagging activity. We show that the statistical
properties of inter-arrival times between subsequent tagging events
cannot be explained without taking into account correlation in users'
behaviors. This shows that social interaction in collaborative tagging
communities shapes the evolution of folksonomies. A consensus
formation process involving the usage of a small number of tags for a
given resources is observed through a numerical and analytical
analysis of some well-known folksonomy datasets.
\end{abstract}




\section{Introduction}

The science of online social networks has recently become a
interdisciplinary research field, since the technological environment
and the number of interacting agents requires the contribution of
researchers such as computer scientists, physicists and
sociologists. A particular example of such social systems are
folksonomies \cite{mika05,furnas06,wu06a, wu06b}, i.e. online
communities of users who, interacting through the World Wide Web,
collaboratively build large and public knowledge bases of discrete
resources such as bookmarks, scientific papers and digital
images. Moreover, folksonomy users participate also in the
classification of individual resources, by labeling each of them with
arbitrarily chosen tags, that is, a (typically small) number of
keywords describing each resource.

Folksonomies act both as public sources of information and as a
storage system for single users, who selfishly collect resources for
their own private use. These two tasks may push the evolution of these
systems in opposite directions \cite{rader06}. As regards the first
purpose, the development of cooperative behavior among users is
crucial. Users have to agree on tag semantics, so that the tag-based
classification of resources be coherent and readable. But, on the
other hand, the popularity of such communities depends on the small
effort demanded to users in the addition of elementary information
units, whose description by tags, though simple, is very approximated
\cite{benkler02}. Besides, the cultural background, the effort and the
needs of users vary a lot throughout the community. This often
generates ambiguous, incomplete or incoherent descriptions of the
collected information and affects the whole accessibility of it.

A fundamental mechanism of consensus building among users is
imitation. For example, consensus triggers the adoption of a given tag
by many users when describing a resource or a whole set of resources,
for descriptive or even strategic purposes. So, the social patterns
of users' interaction reflect onto the statistical distribution of
tags' usage. A highly skewed distribution in the usage of tags has
already been observed, showing that their occurrences vary over many
magnitudes \cite{cattuto06,golder06,shen05}. This reminds the Zipf law
observed in written texts \cite{zipf49}, where the occurrence of words
is distributed according to a power-law. So, the skewness of the tag
frequency distribution may be generated by endogenous mechanisms, or
alternatively be the results of the statistical properties of the
underlying language.

Less attention has been devoted so far to the statistics of tag
dynamics. The growth of the vocabulary, i.e. the number of distinct
tags as a function of time, has been empirically discovered to be
sub-linear in different social tagging systems, and appropriate models
have been developed to reproduce such growth rate, along with the
frequency distribution of tags \cite{cattuto06, halpin07}.

Frequent and rare tags, of course, occur with different inter-arrival
times, but a clear picture of correlations of the same tags by
different users has not been drawn so far. In the following, we will
study the statistics of inter-arrival times in some well-known
collaborative tagging system, where the large number of users allows a
reliable empirical analysis, and will try to find evidences in favor
or against the presence of correlation and collaboration patterns
through the detection of regularities in the temporal statistics of
tags arrival.

Similar analysis have already been performed for other data sets,
namely texts, showing that the distribution of word occurrence is not
random and deviation from a Poissonian picture are present. Fat tails
in the distribution of word inter-arrival times have been detected in
texts, and put into relation with the underlying semantics
\cite{church95,katz96,sarkar05,alvarez06,altmann09}. We will focus here on a
different kind of word sequence, that is, the sequence of tags used by
annotating users in some web-based social tagging community, to
describe the relative resources.

\section{Datasets}

The datasets studied here describe the tagging activities in some
well-known collaborative bookmarking websites including del.icio.us,
Bibsonomy and CiteULike. The data reports individual tag assignments
posted by users in chronological order. Each tag assignment is a
triplet formed by a user, a resource and a tag. Resources are URL in
del.icio.us, while Bibsonomy and CiteULike collect scientific
citations. Tags are keywords associated by users to describe
resources. Each user can assign an arbitrary number of tags to the
same resource in a single post, so more than one tag assignments may
come at the same time.

Such datasets comprises 140306315 tag assignments, with 2482873 tags
for 18778597 resources for the bookmarking website del.icio.us. The
The CiteULike dataset collects 571340 tag assignments with  199512
resources and 51080 tags. The Bibsonomy dataset includes 671808 tag
assignments, with 206942 resources and 58756 tags.

\section{Tag dynamics: observations}
\label{sec:semi-dynam-folks}

Correlations in the behavior of user collaborating in tagging
resources online can be studied by inspecting the temporal statistics
of tag usage. Time, here, is discrete e is measured in number of
successive posts. For example, one can study the inter-arrival time of
tags, that is, the time length occurring between two subsequent tag
assignments involving the same tag. If users behave independently,
tags are added with a constant probability at each time
unit. Accordingly, the arrival of tag would be described by a
Poissonian process, where each occurrence is uncorrelated from the
previous one. In this case, inter-arrival times are distributed
according to an exponential distribution with a well-defined average
inter-arrival times given by $1/f$, where $f$ is the tag frequency
\cite{cattuto07}.

\begin{figure}
  \centering
%  \resizebox{\textwidth}{0.8\textheight}{
\includegraphics[width=0.45\textwidth]{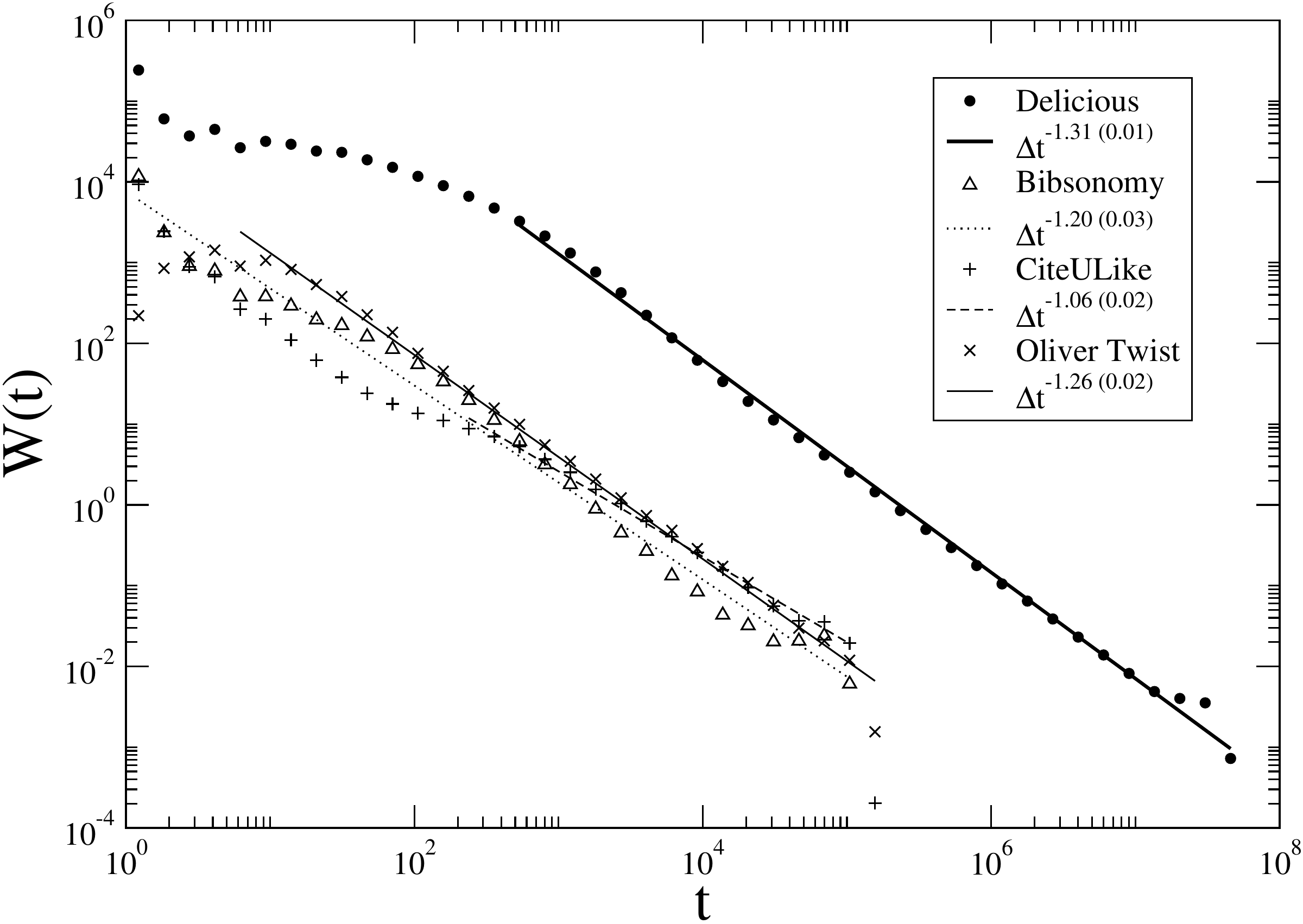}
  \caption{Tag inter-arrival times distribution $W(t)$ in collaborative
    tagging communities as a function of $t$}
  \label{intertimedistribution}
\end{figure}

By contrast, observed individual tag inter-arrival times distribution
shows that inter-arrival times span over all time scales, with a
fat-tailed distribution, as shown in figure
\ref{intertimedistribution}. The number of inter-arrival times of
time length $t$, computed over all tags, is a power law $W(t) \propto
t^{-\gamma}$, with $\gamma \simeq 1.3$ in different tagging systems.

The latter analysis has been performed on a subset of ``stationary''
tags, that is, tags that occurs throughout the whole datasets. This
aims to exclude tags that start or stop occurring in the dataset
during the time window covered by it. These could be frequently
occurring tag with short typical inter-arrival time, though their
observed frequency maybe small because of the partial overlap between
the dataset time window and their lifespan. Thus, a tag with frequency
$f$ is called ``stationary'' if its first occurrence time and the time
interval between its last occurrence and the end of the dataset time
window are both lesser than $1/f$.

\begin{figure}
  \centering
%  \resizebox{\textwidth}{0.8\textheight}{
\includegraphics[width=0.45\textwidth]{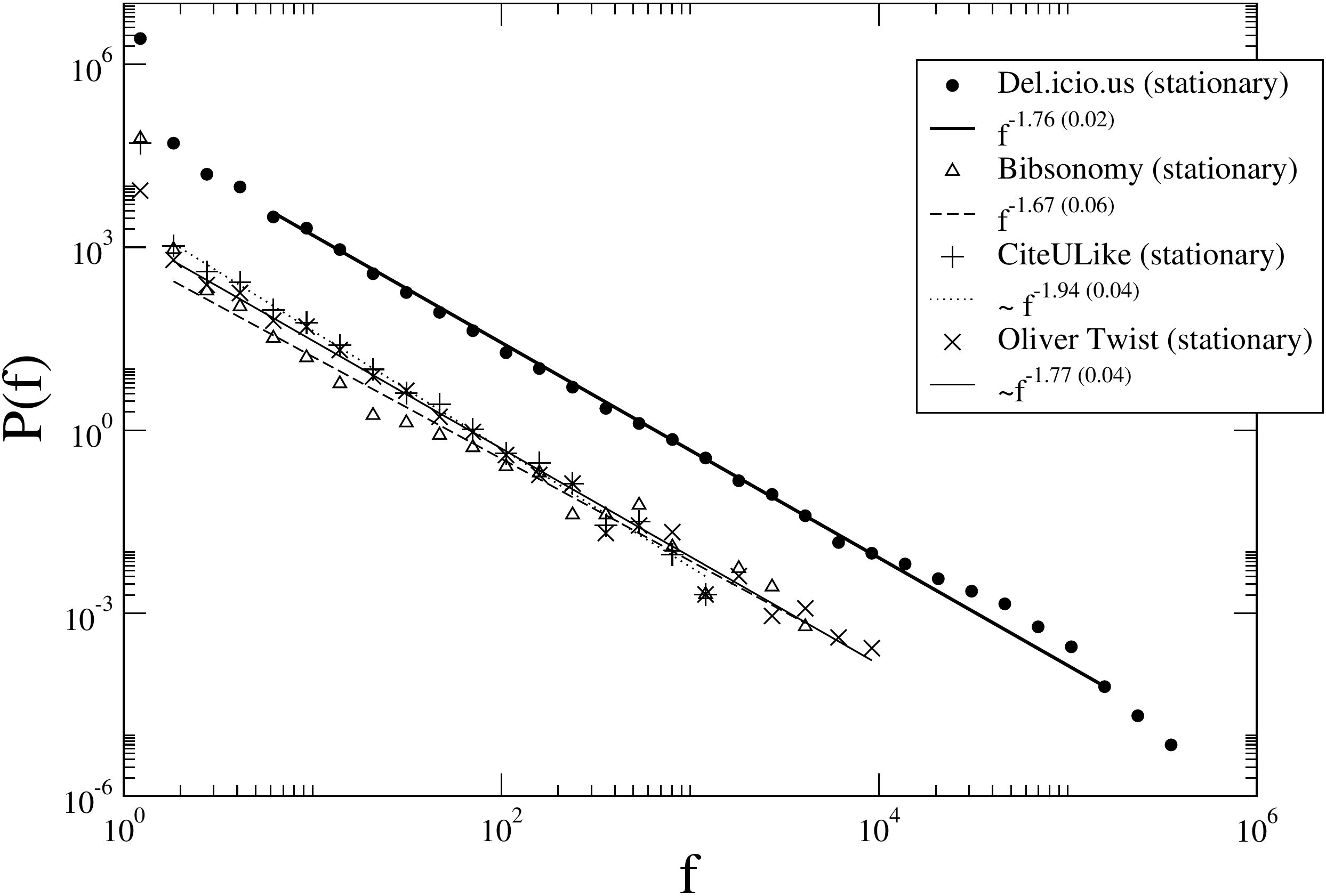}
  \caption{Stationary tag occurrence distribution in collaborative
  tagging communities.}
  \label{zipf}
\end{figure}

Nevertheless, this power--law behavior maybe the consequence of the
uneven distribution of tag occurrences, which is known to follow a
Zipf law. As reported in figure \ref{zipf}, the number of tags
occurring $f$ times is a power--law $P(f) \propto f^{-\beta}$ in
several collaborative tagging communities, with $\beta \simeq
1.7$.

\begin{figure}
  \centering
%  \resizebox{\textwidth}{0.8\textheight}{
\includegraphics[width=0.45\textwidth]{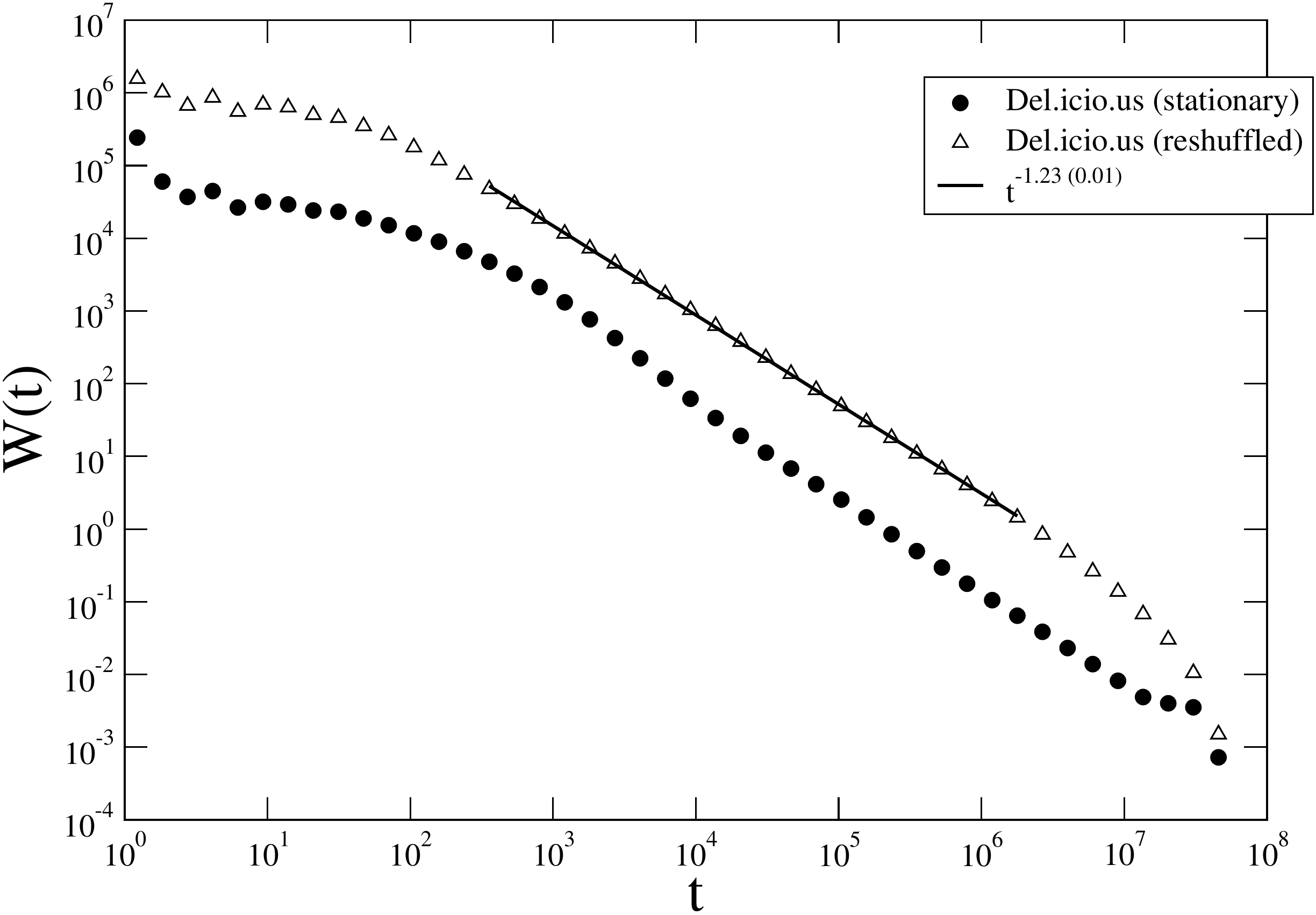}
  \caption{Comparison between the tag inter-arrival times distribution
  in the original Del.icio.us dataset (stationary tags) and in an
  artificial one where the time-ordering of tag assignments has been
  randomly reshuffled.}
  \label{reshuffling}
\end{figure}

The fat tails in figure \ref{intertimedistribution} may be determined
by the large number of tags with low frequency (i.e., long
inter-arrival times). To verify this, one reshuffles the time ordering
of tags by reassigning them to randomly chosen posts. This way, time
correlations are removed and the distribution of inter-times is
determined solely by the Zipf law in the frequency of tags. From now
on, we limit our statistical analysis to the larger Del.icio.us
dataset, where richer data allow a more reliable statistical
analysis. However, the numerical and analytical results presented here
hold within a reasonable approximation in other social bookmarking
communities.  As checked in figure \ref{reshuffling}, the inter-arrival
time distribution changes slightly from the power--law behavior
described above. Therefore, the distribution $W(t)$ is no signature of
complex correlation patterns. This can also be easily understood by a
simple analytical argument, shown in the next section. 

Thus, one should observe individual tag inter-arrival time
distribution. which, of course, display a poorer statistics. Here one
finds different patterns for high--frequency tags and low--frequency
tags.  The first display a fast decay in the distribution for large
values of the inter-arrival time. Reasonably, tags that occur less
frequently display longer inter-arrival times with a finite
probability. Their inter-arrival times distribution decay as a power
law for large values of $\Delta t$.

\begin{figure}
  \centering
%  \resizebox{\textwidth}{0.8\textheight}{
\includegraphics[width=0.45\textwidth]{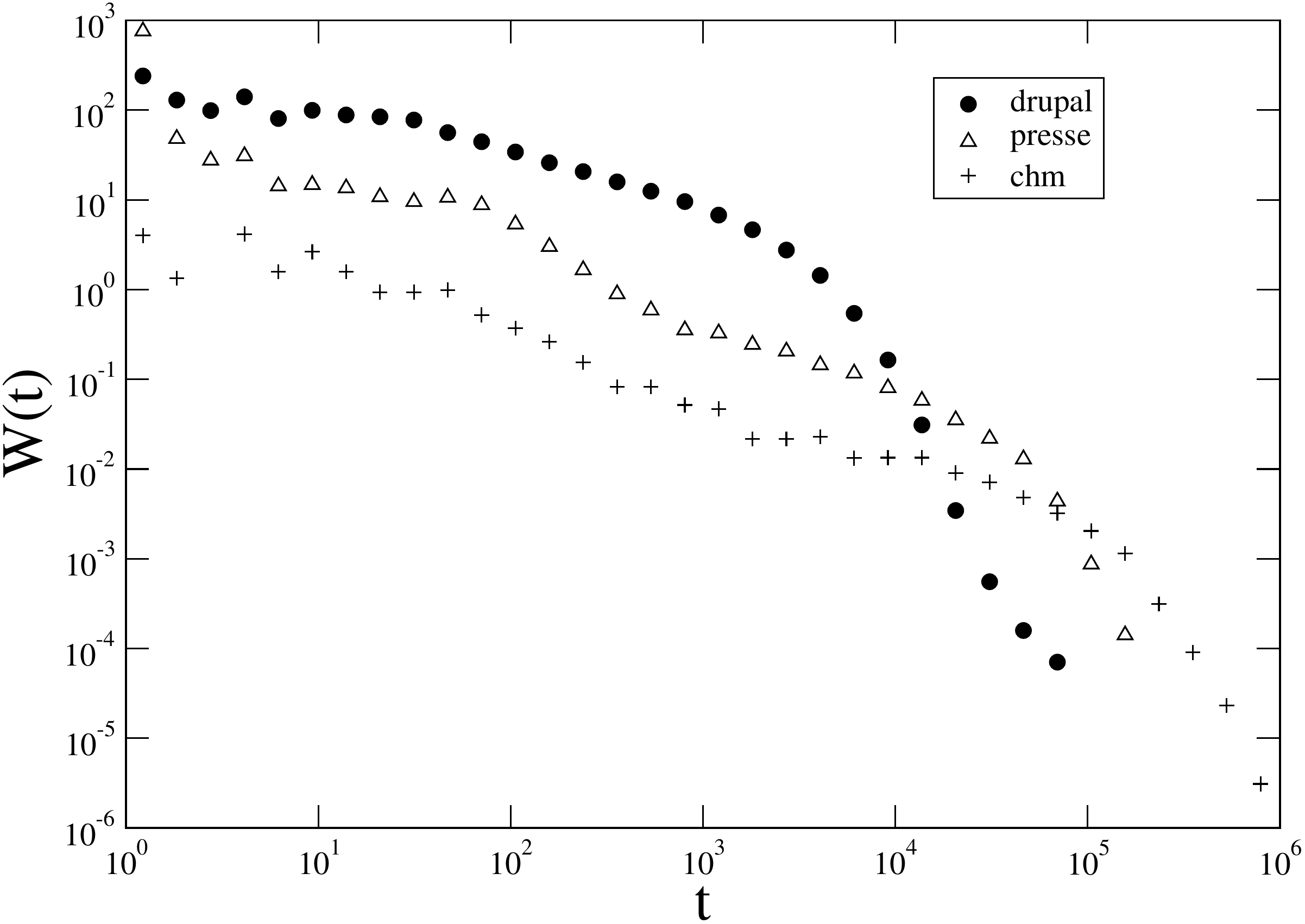}
  \caption{The inter-arrival times distribution for individual tags
    ``drupal'' (33442 occurrences), ``presse'' (5011) and ``chm''
    (999) in the collaborative tagging system del.icio.us.}
  \label{singletags}
\end{figure}

\begin{figure}
  \centering
%  \resizebox{\textwidth}{0.8\textheight}{
\includegraphics[width=0.45\textwidth]{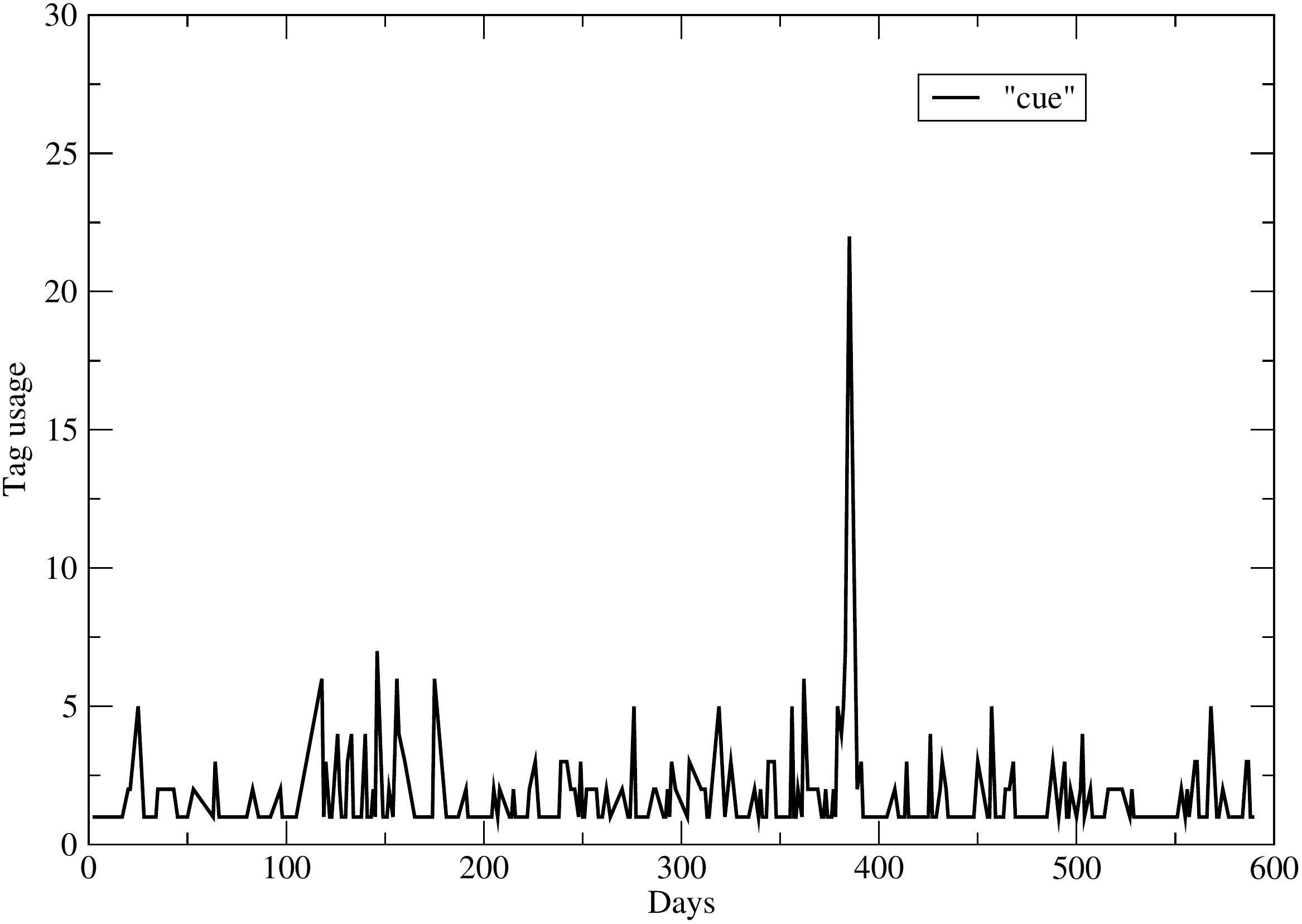}
  \caption{The weekly usage of the tag ``cue'' during the two years
    covered by the del.icio.us dataset, displaying periods of high
    activity and sharp activity peaks. The x-axis reports the number
    of weeks since 1st January 1970}
  \label{tagburst}
\end{figure}

The presence of power laws in the distribution of inter-arrival times
is often put in strict relation with processes taking place in
``avalanches'', i.e. with long period of stability with sudden bursts
of activity of all scales of magnitudes, limited only by finite size
effects, as shown in an example reported in figure
\ref{tagburst}. Scale-invariance in the distribution of inter-time
distribution corresponds to unpredictability of future events, given
the past time series \cite{jensen98}.

By reshuffling tags, time correlations would be removed, and the
curves corresponding to those plotted in figure \ref{reshuffling}
would exhibit an exponential decay

\section{Inter-arrival times distribution}

The statistics of inter-arrival times for high-frequency and
low-frequency tags can be simply related.  The inter-arrival time
distributions of individual tags can be modeled by a scaling function
depending on the tag frequency $f$, written as
\begin{equation} \label{wf_def}
W_f(t) \propto R(f) t^{-\alpha} g(f^\psi t)
\end{equation}
where $W_f(t)$ is the number of inter-arrival times of time length $t$
for a tag of frequency $f+1$, and $g$ is a function which is constant
for low values of the argument and decays rapidly after a given
cut-off value, i.e. $g(x) = 1$ for $x\ll 1$ and $g(x)=0$ for $x \gg 1$.
Since the total number of inter-times for a tag with frequency $f+1$ is
$f$, $W_f(t)$ is normalized by
\begin{equation}
\int_0^\infty W_f(t) dt = f.
\end{equation}
By using the definition \ref{wf_def}, this leads to 
\begin{equation}
f \propto R(f)f^{\psi(\alpha-1)}\int_0^\infty x^{-\alpha}g(x)dx,
\end{equation}
where $x=f^{\psi}t$, so that $R(f) = f^{1+\psi (1-\alpha)}$.

\begin{figure}
  \centering
%  \resizebox{\textwidth}{0.8\textheight}{
\includegraphics[width=0.45\textwidth]{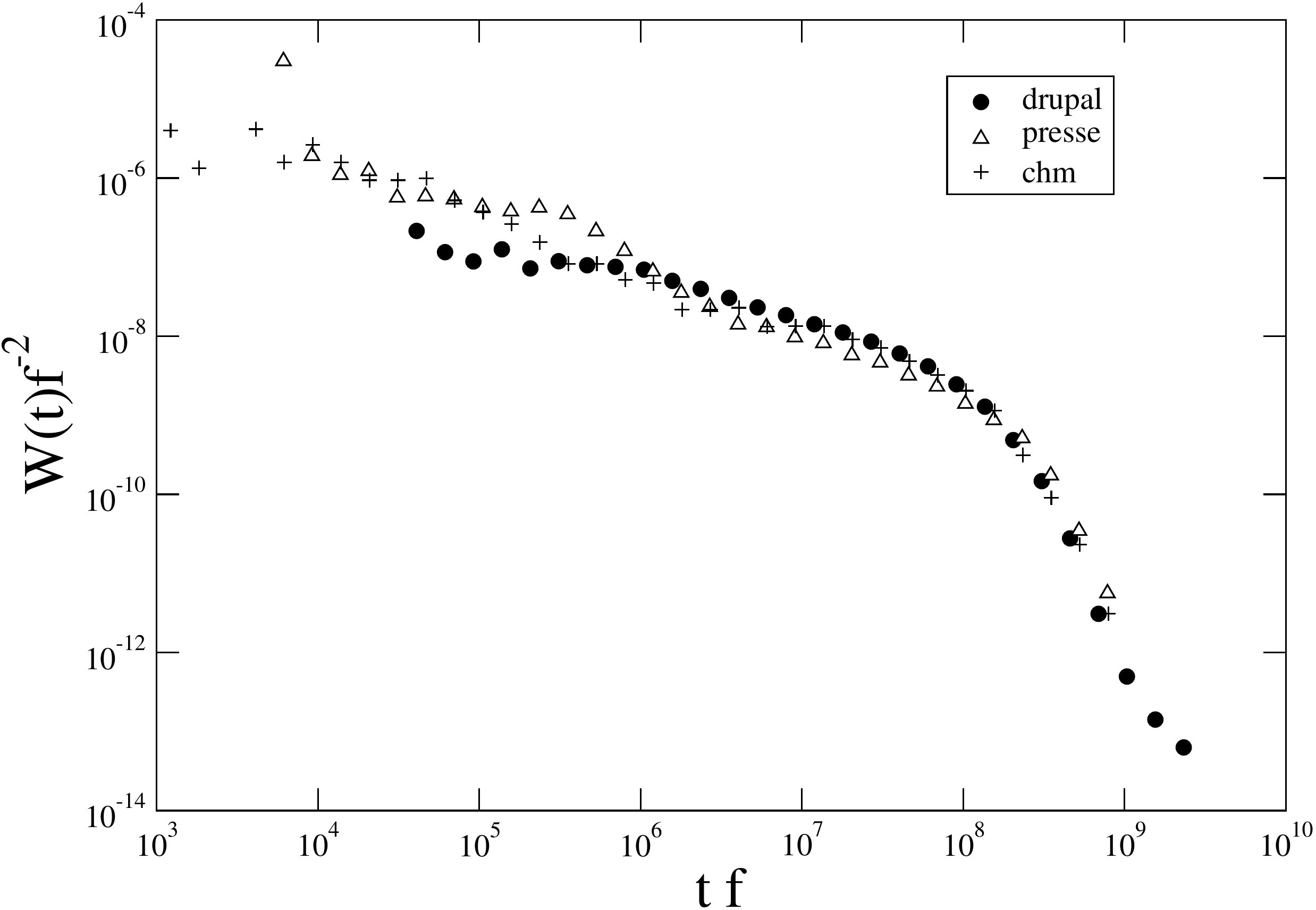}
  \caption{ The cutoff function $g(ft) =
    \frac{W_f(t)}{f^{2-\alpha}t^{-\alpha}}$ plotted against $ft$ for
    the inter-arrival times distribution of individual tags ``drupal''
    (33442 occurrences), ``presse'' (5011) and ``chm'' (999) in the
    collaborative tagging system del.icio.us.}
  \label{cutoff}
\end{figure}

The inter-arrival times distribution observed over all tags $W(t)
\propto t^{-\gamma}$, which receives contribution by the occurrence
of tags of all frequencies distributed according to the law \ref{zipf},
can be written as
\begin{equation}
W(t) = \int_0^\infty P(f)W_f(t) df,
\end{equation}
which, after replacing $P$ and $W_f$ by their functional form, reads
\begin{equation}
W(t) \propto t^{\frac{\beta-2}{\psi} - 1} \int_0^\infty
x^{\beta(1-\alpha)-1+\frac{1}{\psi}}g(x)dx.
\end{equation}.
Thus, one obtains the relation $\psi = \frac{2-\beta}{\gamma-1}$. By
replacing the observed values for $\beta$ and $\gamma$, the relation
yields $\psi \simeq 1$ for del.icio.us, $W(t) \propto t^{\beta-3}$ and 
\begin{equation} \label{wft}
W_f(t) \propto f^{2-\alpha} t^{-\alpha} g(tf) 
\end{equation}.  
The value of the exponent $\alpha \simeq 0.75$, measured by the
inter-arrival times distribution in del.icio.us, is verified in the
figure \ref{cutoff}, where the inter-arrival times distributions for
tags with different frequencies collapse on the same function.

After reshuffling tag order as described above, since the arrival of
tags is now a Poissonian process, the distribution \ref{wft} changes
into an exponential function, with inter-arrival times statistics equal
to $W_f^{(P)}(t) \propto f\lambda(f) e^{\lambda(f)t}$. $\lambda(f)$ is
the average inter-arrival times, equal to $f/T$ where $T$ is the time
length of the observed period. The distribution $W(t)$ can thus be
computed as
\begin{equation}
W(t) = \int_0^\infty P(f) f \lambda (f)e^{-\lambda(f)t}df
\end{equation}
that, by replacing the power-law form of $P(f)$ yields $W(t) \propto
t^{\beta-3}$, showing that the reshuffling changes only the single-tag
inter-arrival times distribution but leaves unchanged the overall
inter-arrival times distribution. Therefore, the power-law behavior
observed for $W(t)$ depends only on the frequency skewed distribution
and cannot be used to study the dynamical properties of tag arrival.

\section{Collaborative patterns}

\begin{figure}
  \centering
%  \resizebox{\textwidth}{0.8\textheight}{
\includegraphics[width=0.45\textwidth]{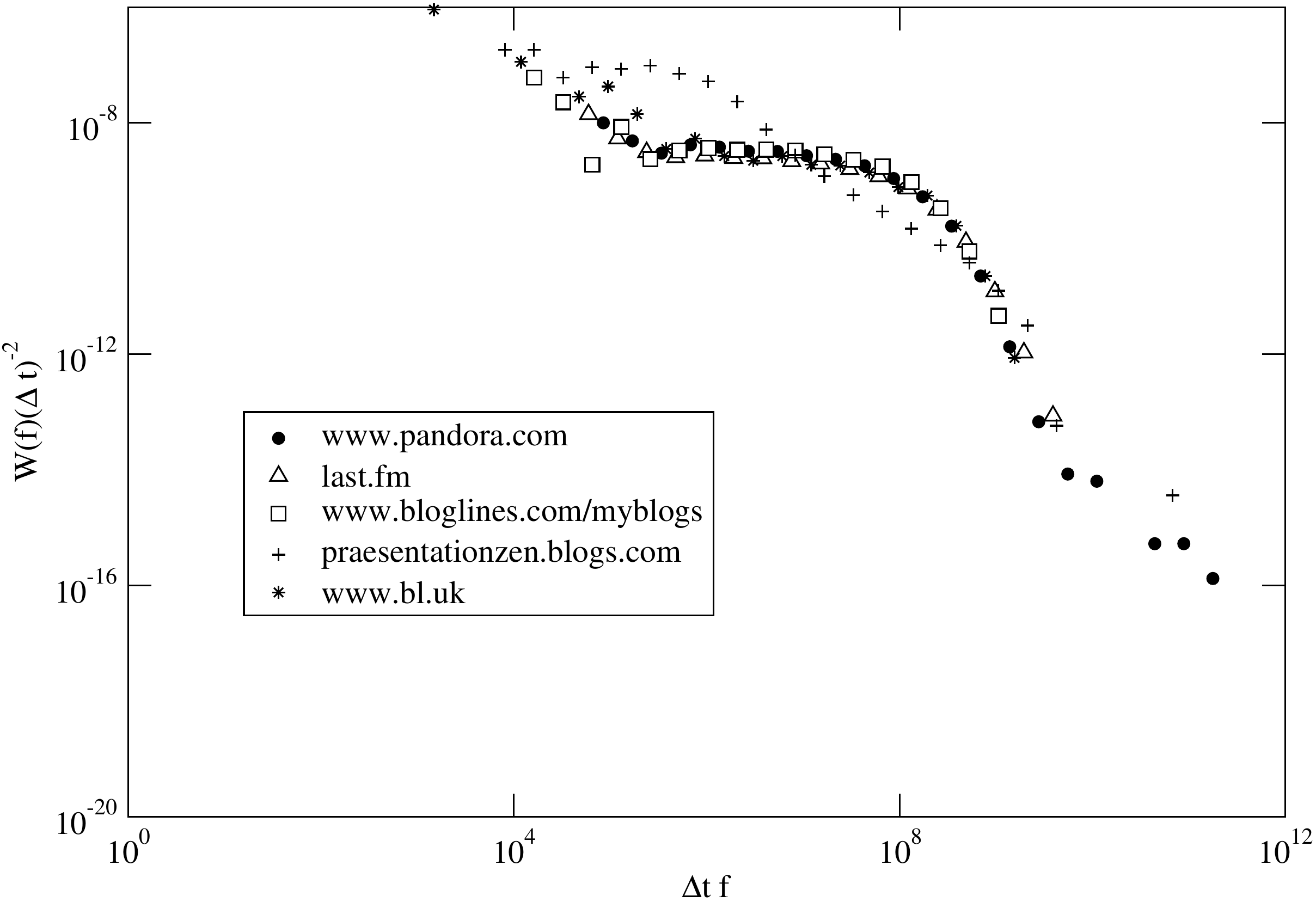}
  \caption{The distribution of inter-arrival times between subsequent
    tag assignments involving the same resource.}
  \label{resintertime}
\end{figure}

The bursty behavior of tagging activities is not in itself a
signature that complexity arises due to the interaction of users. A
clearer sign of user cooperation can be found by analyzing the
temporal pattern corresponding to individual resources. Inter-Arrival
times $t$ between subsequent tagging of the same resource are
distributed according to a power--law with a sharp cut--off for large
values of $t$ going to infinity for less tagged resources, as
displayed in figure \ref{resintertime}. Since a user cannot tag a
resource twice, the fact that individual resource are tagged in
``avalanches'' depends on the contribution of many users. By
contrast, if users were tagging independently one from each other,
$t$ should be distributed as an exponential random variable, as
happens for Poissonian processes. The individual resource inter-arrival
distribution can be analyzed as done above for tags, showing that
resources are tagged in bursts spanning all time length scales.

\begin{figure}
  \centering
%  \resizebox{\textwidth}{0.8\textheight}{
\includegraphics[width=0.45\textwidth]{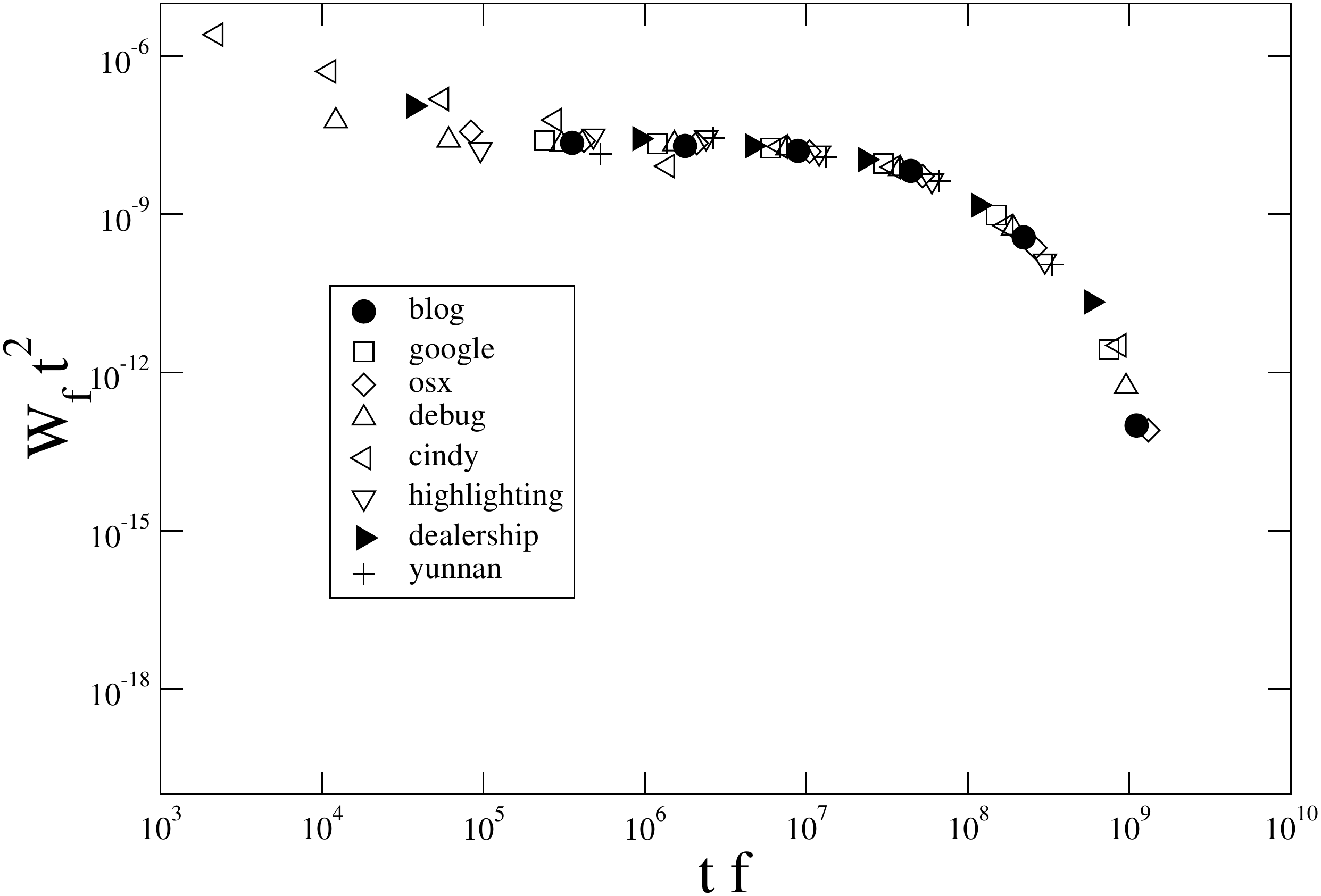}
  \caption{The distribution of inter-arrival times between the first
    usage of some tags by each user, divided by the number of such events
    $f_1$, plotted as a function of $t_1f_1$.}
  \label{wft1}
\end{figure}

A similar avalanche-like pattern can be observed by taking into
account only the first usage of a tag by each user, and observing the
inter-arrival times distribution $W_{f_{1}}(t_1)$ of this special tagging
events, where $t_1$ refers to their time separation and $f_1$ is the
number of such events. This way, one removes the possibility that the
the short inter-arrival times are originated by users who often use a
given tag for their own interests, and long inter-arrival times may
come by the numerous users who seldom tag resources with that
particular tag. If this was the case, the skewed distribution would
just be the result of the superposition of heterogeneous, yet
independent, usage patterns. Interestingly, the distribution of
inter-arrival times of a given tag, when one limits the observation to
the first usage of that tag by each user, follows the same statistics
observed above when one takes into account the whole tagging
activity. In particular, the relation reported in eq. \ref{wft} holds
also for the inter-arrival times $t_1$, as shows the collapse reported
in figure \ref{wft1}. If relation \ref{wft} holds, tags are
``discovered'' by users in a correlated and bursty manner.

However, this is not yet a proof of cooperation among online users. In
fact, bursts of attention may arise by both a direct mutual influence
between users one on each other; otherwise, users may independently be
influenced by the same sources of information and news, where
attention bursts may originate without any interaction among them.

The stream of tag assignment involving a given resource, though,
carries a clearer evidence of users interaction. By plotting the
number of distinct tags, i.e. the vocabulary, used for a resource as a
function of the number of tag assignments to it, one observes a
sub-linear vocabulary growth: so, the pace at which new tags are
introduced by users to describe a resource decreases with time, and
new tags are introduced less and lesser. In other words, users tend to
employ the same tags used by previous peers when describing the same
resources.

\begin{figure}
  \centering
%  \resizebox{0.5\textwidth}{0.25\textheight}
\includegraphics[width=0.45\textwidth]{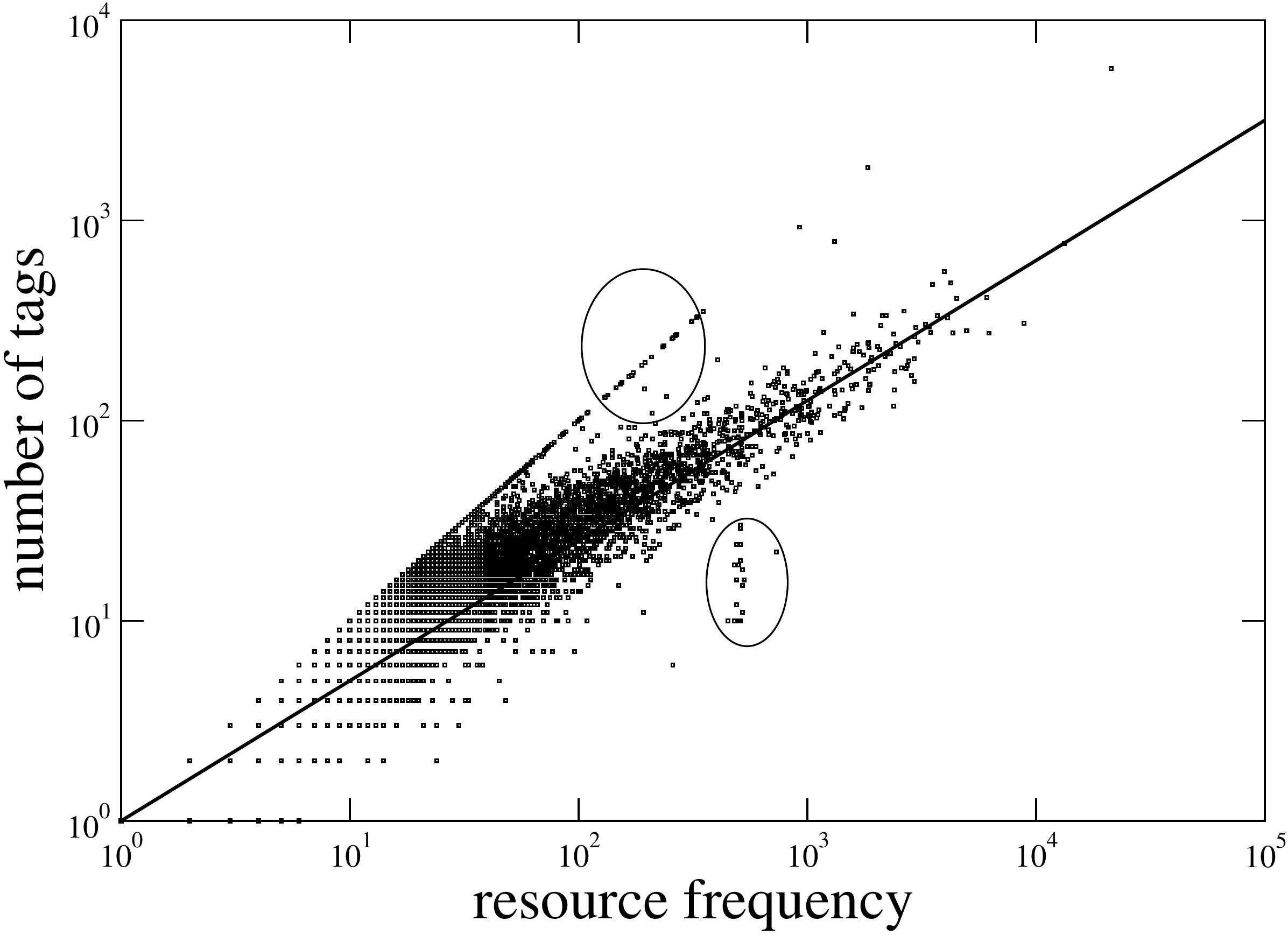}
  \caption{The number of distinct tags $n(f)$ assigned to a resource
    as a function of the number $f$ of tag assignments involving the
    resource. The majority of resources lie approximately on the line
    $n(f) \propto t^{0.7}$. The plot shows also two clusters of
    resources (in the ellipses) lying away from the line. A direct
    inspection reveal that that resources are malicious ``spam''
    bookmarks.}
  \label{tagsvsfreq}
\end{figure}

The figure \ref{tagsvsfreq} shows that the sub-linear relation between
tag assignments and number of distinct tags involving a single
resource holds for the large majority of them. Interestingly, this
relation is not respected by ``spam'' bookmarks, that is, by tag
assignments violating of the collective agreement about tag semantical
organization. As other signatures of complex features, so, this
relation may reveal useful in methods of spam detection
\cite{cattuto07,capocci08}.

A deeper insight into the collective development of a tag vocabulary
associated to a resource is provided by studying the Inverse
Participation Ratio ($IPR$) of tags in such vocabulary.  Let
${v_i}_{i=1,...,N}$ be the components of a vector {\bf v}, such that
$\sum_{i=1,N}v_i^2=1$. The definition of $IPR$ is $IPR =
\sum_{i=1,N}v_i^4$. If all components are equal to $v_i = 1/\sqrt{N}$,
$IPR$ is equal to $1/N$. Conversely, if all components are null but
one, $IPR = 1$. So, the IPR describes the number of a vector
components that contribute significantly to the vector
norm. Analogously, the $IPR$ of tag streams computed on the relative
frequency of tags represents the number of significant tag used to
describe a given resource.

\begin{figure}
  \centering
%  \resizebox{\textwidth}{0.8\textheight}{
\includegraphics[width=0.5\textwidth]{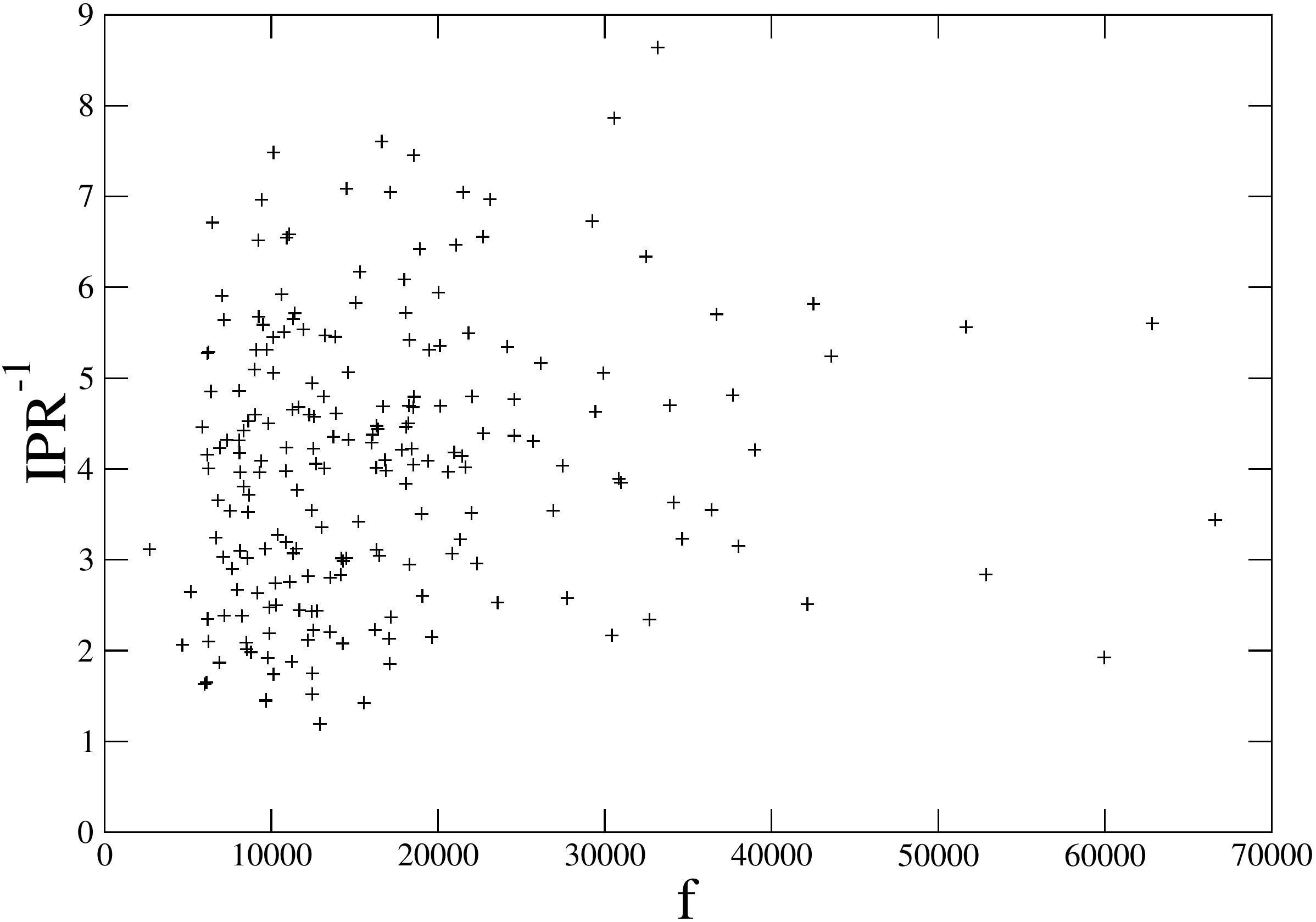}
  \caption{
    The $IPR$ of tag streams associated to individual resources as a
    function of the number of tag assignments $f$ involving them.
}
  \label{ipr}
\end{figure}

As shown in figure \ref{ipr}, the number of significant tags is rather
constant even for resources tagged thousands of times, showing that a
consensus is reached among users about how to describe a given
resource and following users do not add new significant tag but rather
employ already used ones.

\section{conclusions}

We have studied empirically and analytically the behavior of users in
some well-known collaborative tagging online systems, where a large
number of users collects resources and classify them by attaching a
number of labels, called tags, to each of them. A resource can be
tagged by many users, and thus be tagged by a large number of labels.

We have analyzed the statistics of inter-arrival times of tags,
i.e. the time interval occurring between two subsequent occurrences of
a same tag, and of resources. We have uncovered non-trivial
statistical properties, which can be related to avalanches in the
tagging activities. Such bursty behavior shows that the tagging
activity by different users is strongly correlated. Regularities in
the inter-arrival times distribution are studied analytically, so that
the dynamics of rare and frequent tags can be unified by a unique law,
which depends only on the frequency parameter $f$.  Moreover, we have
shown that users of tagging systems find a consensus about the tag
description of each resource. In fact, we have empirically shown that
the number of significant tags for each resource is rather constant,
even for resources that have been tagged by thousand of heterogeneous
users. A by-product of our analysis regards the detection of spam in
such freely accessible communities. The number of distinct tags
attached to a resource, i.e. the resource vocabulary length, grows
sub-linearly with the number of tagging events involving that
particular resource, with a relation which holds with good precision
for a large majority of tags. Two well-defined subset of tags,
however, do not satisfy such relationship between the resource
occurrence and the resource vocabulary length. A direct inspection of
such tags reveals that the latter have been added during malicious
spam activity. This suggests a fast method to detect spam in
collaborative tagging systems.

\section{Acknowledgments}
This research was supported by the TAGora project
(FP6-IST5-34721) funded by the Future and Emerging Technologies
program (IST-FET) of the European Commission.  We thank C. Cattuto for
many stimulating discussions.

\bibliographystyle{abbrv}
\bibliography{paper}

\end{document}